\documentclass[journal]{IEEEtran}      

\pdfoutput=1
\IEEEoverridecommandlockouts           

\usepackage{graphicx}
\usepackage{graphics}
\usepackage{amsmath,bm}
\usepackage{amssymb}
\usepackage{algorithm}
\usepackage{algpseudocode}
\usepackage{cite}
\usepackage{hyperref}
\usepackage[caption=false]{subfig}
\usepackage{booktabs}
\usepackage{siunitx}
\usepackage{float}
\hypersetup{hidelinks}

\algdef{SE}[DOWHILE]{Do}{doWhile}{\algorithmicdo}[1]{\algorithmicwhile\ #1}%

\newcommand{\mathactivatecomma}{%
  \begingroup\lccode`~=`\,
  \lowercase{\endgroup\edef~}{\mathchar\the\mathcode`\,\penalty0 }}

\algnewcommand{\Initialize}[1]{%
  \State \textbf{Initialize: $i \in \mathcal{V}$}
  \Statex \hspace*{\algorithmicindent}\parbox[t]{.8\linewidth}{\raggedright #1}
}

\algnewcommand{\Iteration}[1]{%
  \State \textbf{Iteration $(k\geq 0)$: $i \in \mathcal{V}$}
  \Statex \hspace*{\algorithmicindent}\parbox[t]{.8\linewidth}{\raggedright #1}
}

\algnewcommand{\Output}[1]{
  \State \textbf{Output: $i \in \mathcal{V}$}
  \Statex \hspace*{\algorithmicindent}\parbox[t]{.8\linewidth}{\raggedright #1}
}

\title{\LARGE \bf
An Effective Image Restorer: Denoising and Luminance Adjustment for Low-photon-count Imaging}

\author{Shansi Zhang and Edmund Y. Lam

\thanks{S. Zhang and E.Y. Lam are with the Department of Electrical and Electronic Engineering, The University of Hong Kong, Pokfulam, Hong Kong {\tt\small e-mail: sszhang@eee.hku.hk, elam@eee.hku.hk}.}
}

\begin{document}

\maketitle
\thispagestyle{empty}
\pagestyle{empty}

\begin{abstract}
Imaging under photon-scarce situations introduces challenges to many applications as the captured images are with low signal-to-noise ratio and poor luminance. In this paper, we investigate the raw image restoration under low-photon-count conditions by simulating the imaging of quanta image sensor (QIS). We develop a lightweight framework, which consists of a multi-level pyramid denoising network (MPDNet) and a luminance adjustment (LA) module to achieve separate denoising and luminance enhancement. The main component of our framework is the multi-skip attention residual block (MARB), which integrates multi-scale feature fusion and attention mechanism for better feature representation. Our MPDNet adopts the idea of Laplacian pyramid to learn the small-scale noise map and larger-scale high-frequency details at different levels, and feature extractions are conducted on the multi-scale input images to encode richer contextual information. Our LA module enhances the luminance of the denoised image by estimating its illumination, which can better avoid color distortion. Extensive experimental results have demonstrated that our image restorer can achieve superior performance on the degraded images with various photon levels by suppressing noise and recovering luminance and color effectively.
\end{abstract}


\section{Introduction}\label{sec:introduction}

\begin{figure*}[ht]
\centering
\includegraphics[width=1\textwidth]{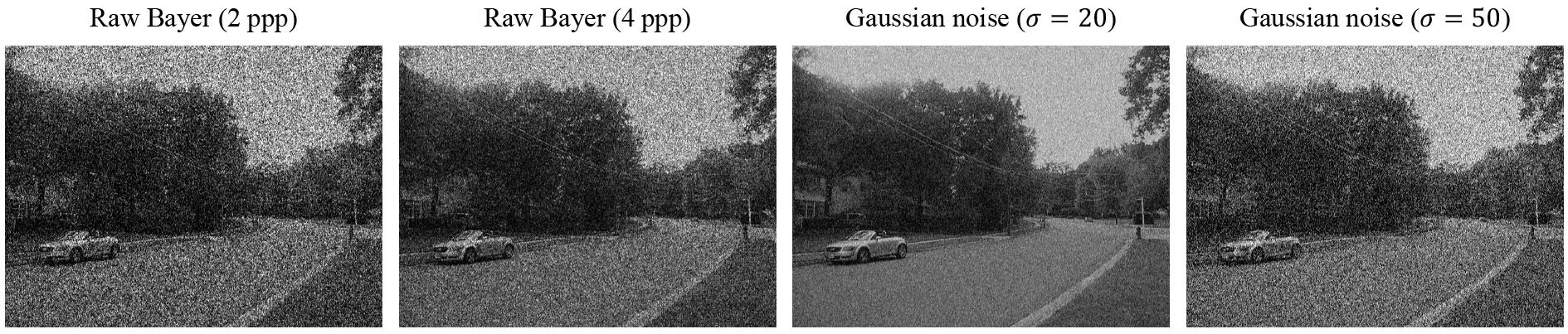} 
\caption{Two raw Bayer images with low photon counts ($2$ ppp and $4$ ppp) and two gray images distorted by additive Gaussian noise with different strengths ($\sigma=20$ and $\sigma=50$).}
\label{fig:low-photon-count}
\end{figure*}

Imaging under photon-scarce situations bring challenges to many applications, such as surveillance, robotics and computational photography. As the photons become scarce, the signal-to-noise ratio of the captured images will be lower. The raw images may be severely corrupted by the noise under extremely low photon levels, which introduces large difficulties to image restoration. To alleviate the problem of photon-scarce imaging, some advanced sensors, including single-photon avalanche diode (SPAD) and quanta image sensor (QIS), which has single-photon detection sensitivity and photon-counting capability\cite{Elgendy2018}, are developed. These sensors are good candidates for low-light imaging due to their high sensitivity to photons. 

QIS is a new type of image sensor, where each pixel is partitioned into many smaller pixels, called `jots', and each jot is a single-photon detector\cite{Ma2015,Chi2020}. Although QIS has high sensitivity to photons, its captured raw images are also with low signal-to-noise ratio under low-photon-count conditions. There are some researches\cite{Abhiram2019,Chi2020,Gnanasambandam2020,Elgendy2021} focusing on the image reconstruction with QIS. Here, we treat QIS as a generic concept representing the photon-counting property so that the captured images can record the average number of photons that each pixel detects during the exposure period. Then, we use the `photons per pixel' (ppp) as a rough measurement for the degradation degree of each raw image. As the average photon count decreases, the raw image is more severely corrupted. Thus, we study on the raw image restoration under low-photon-count imaging, where the Poisson-Gaussian mixed noise dominates the images. Fig.~\ref{fig:low-photon-count} shows two raw Bayer images with low photon counts ($2$ ppp and $4$ ppp) that we will deal with in this paper, as well as two gray images distorted by additive Gaussian noise with different strengths ($\sigma=20$ and $\sigma=50$). As can be seen, the raw images under photon-scarce conditions are even worse than the `heavy noise' situation in the common denoising literatures, which reflects the challenge of our task.

The raw images with low photon counts suffer from severe noise and poor luminance, which can not be dealt with effectively by the previous methods. Considering the severity of degradations, we restore each image through two steps, with separate denoising and luminance adjustment. The image is first denoised by our multi-level pyramid denoising network (MPDNet), which can suppress noise for the images with various photon levels. Then, the denoised image is further enhanced by our luminance adjustment (LA) module, which can adjust the luminance adaptively according to the estimated illumination of the input image, so that we can obtain more visually compelling image. In addition, our framework is a generic solution to the degraded images with noise and unsuitable luminance, not only limited to the low-photon-count imaging. The main contributions of this paper are summarized as follows:
\begin{itemize}
\item We design a multi-skip attention residual block (MARB), which integrates multi-scale feature fusion and attention mechanism to obtain better feature representation. Moreover, it has much fewer parameters than the common residual block.
\item We develop a MPDNet, which adopts the idea of Laplace pyramid to learn the small-scale noise map and larger-scale high-frequency maps at different levels. The learned high-frequency components are gradually added to the up-sampled denoised images to obtain the clean images with proper sharpness. We also introduce multi-scale input image feature extraction to encode richer contextual information. 
\item Then, we design an LA module, which enhances the luminance of the denoised image according to its estimated illumination. Our LA module can better avoid color distortion by fully reserving the color information of the input image.
\item Our framework for image restoration is very lightweight and efficient. Extensive experimental results demonstrate its effectiveness and robustness for restoring images with severe noise and poor luminance.
\end{itemize}

\section{Related work}\label{sec:related work}

\subsection{Image denoising}
The techniques for image denoising mainly include model-based and deep learning-based methods. The model-based methods utilize the priors of image or noise to obtain clean images through traditional optimization algorithms. The concrete methods include non-local filter\cite{Buades2005,Dabov2007,Sun2014,Li2016}, sparse coding\cite{Elad2006,Mairal2009,Xu2018}, external priors\cite{Zoran2011,Chen2015} and low rank\cite{Dong2013,Gu2014}. However, these methods only assume additive white Gaussian noise (AWGN) and are not applicable to more complicated noise. 

Currently, most researches on image denoising are based on deep learning. Some methods\cite{Chen2017,Lefkimmiatis_2017} combine prior knowledge and deep neural network, which may not be robust enough due to the limitation of priors. Some methods adopt completely data-driven approach to learn the noisy-to-clean mapping. Zhang et al.\cite{Zhang2017} proposed DnCNN, which uses residual learning to remove Gaussian noise with various noise levels. Guo et al.\cite{Guo2019} proposed CBDNet by considering realistic noises for blind denoising, with a noise estimation sub-network to avoid under-estimation for noise level by asymmetric learning. Hong et al.\cite{Hong2019} divided the denoising task into several local subtasks, each of which was addressed by a network trained on its local space, and these subtasks were combined by mixture weights during testing. Anwar et al.\cite{Anwar2019} developed RIDNet for real image denoising, which uses residual on the residual structure to ease the low-frequency flow with feature attention for exploring the channel dependencies. Tian et al.\cite{Tian2020} proposed ADNet, with global and local information integration and attention mechanism to achieve real and blind denoising. Zamir et al.\cite{Zamir2020} proposed MIRNet for image restoration, which contains multi-resolution branches with information exchange, spatial and channel attention mechanism and attention-based feature fusion to learn enriched features.

\subsection{Low-light image enhancement}
The existing methods for low-light image enhancement can be mainly divided into two types: direct mapping from the low-light image to the normal-light image and Retinex-based decomposition and enhancement. 

For the direct mapping approach, Lore et al.\cite{Lore2017} proposed LLNet, with an encoder-decoder structure to adaptively brighten the low-light images. Lv et al.\cite{Lv2018} proposed MBLLEN, which contains multiple branches to extract features with different levels and a fusion module for multi-branch feature fusion. Jiang et al.\cite{Jiang2021} developed EnlightenGAN, which uses UNet-based architecture with the input illumination channel as the self-regularized attention map, and can be trained by unpaired images. These methods usually adopt residual learning, which can make the learning easier but still can not avoid color distortion. 

For the Retinex-based approach, Wei et al.\cite{Wei2018} proposed deep Retinex-Net, which consists of a Decom-Net for the reflectance and illumination decomposition, and an Enhance-Net for improving the illumination. The reflectance was denoised by BM3D\cite{Dabov2007} and then multiplied with the enhanced illumination to obtain the final output. Zhang et al.\cite{Zhang2019} developed KinD, which introduces a restoration network besides an illumination adjustment network, to better suppress noise for the low-light reflectance. Zhang et al.\cite{Zhang2021b} proposed a deep learning-based decomposition-enhancement method to restore the light field images under low-light conditions. Usually, the decomposition network is not easy to train, which may cause color distortion for the decomposed reflectance and therefore influence the restoration performance.

\section{Proposed method}\label{sec:method}

\subsection{Overall framework}

\begin{figure}[ht]
\centering
\includegraphics[width=0.48\textwidth]{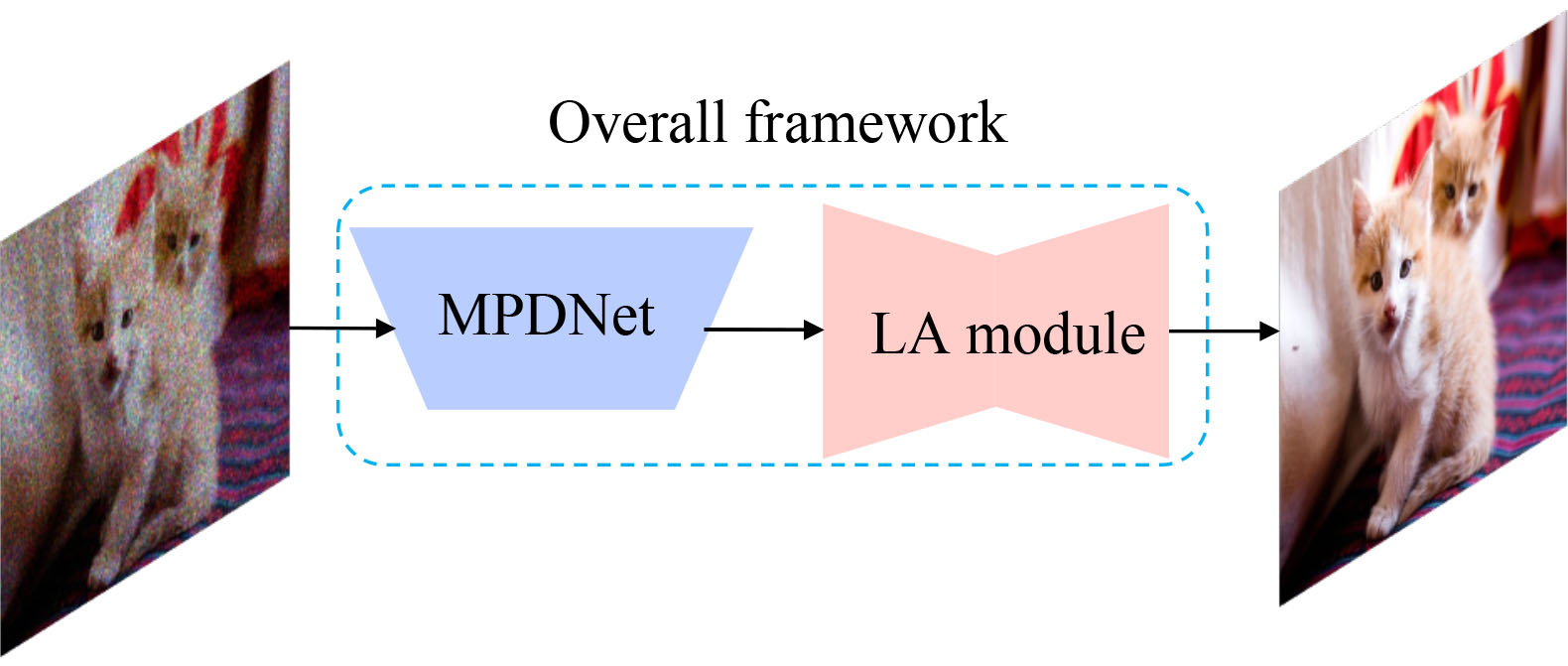} 
\caption{The overall framework consists of a MPDNet for noise suppression and an LA module for luminance adjustment.}
\label{fig:framework}
\end{figure}

Our framework aims to restore the degraded images with severe noise and poor luminance to the visually compelling images. The overall framework consists of a MPDNet for noise suppression and an LA module for luminance adjustment, as depicted in Fig.~\ref{fig:framework}. We design separate networks to achieve different functions, which can reduce the learning difficulty and therefore improve the restoration performance. In what follows, we will provide the detailed architectures of the MPDNet and LA module.

\subsection{Multi-skip attention residual block}

\begin{figure}[ht]
\centering
\includegraphics[width=0.5\textwidth]{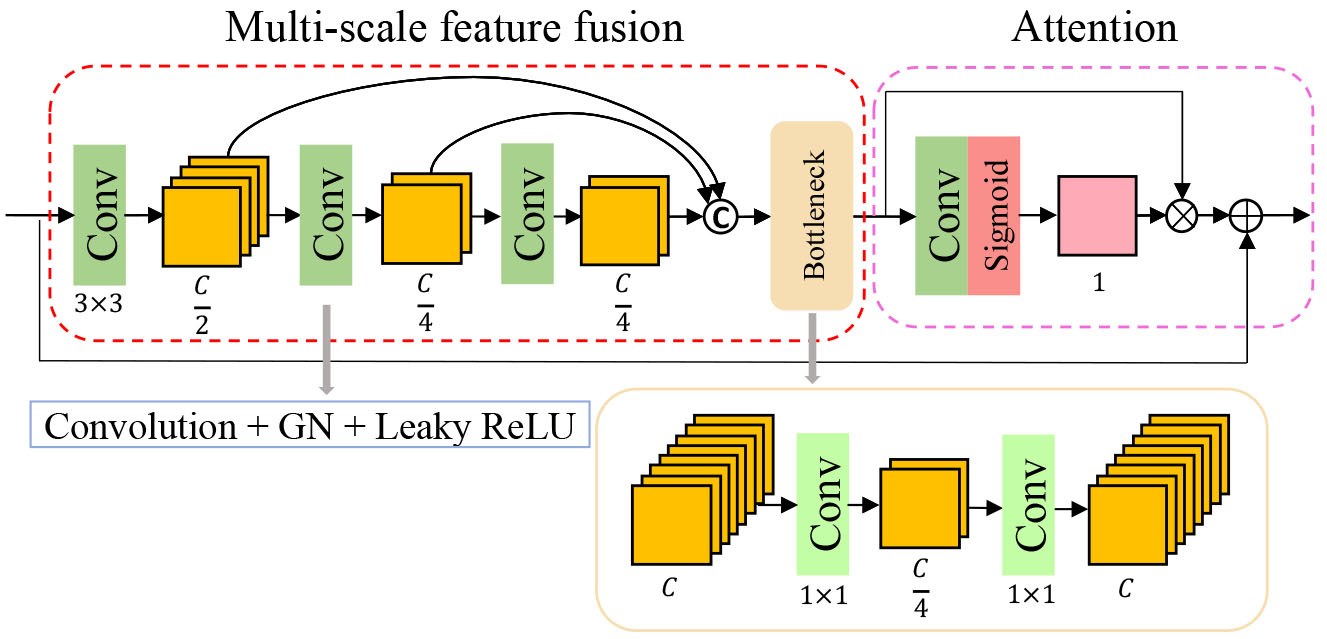} 
\caption{The structure of MARB. The first three convolution layers output $\frac{C}{2}$, $\frac{C}{4}$, and $\frac{C}{4}$ channels, respectively. These outputs are concatenated together, followed by $1\times 1$ convolution layers for feature fusion. Then, the fused features are calibrated by the spatial attention to extract more informative features.}
\label{fig:MARB}
\end{figure}

The main component of our MPDNet and LA module is the multi-skip attention residual block (MARB), as shown in Fig.~\ref{fig:MARB}. First, there are three $3 \times 3$ convolution layers, each of which includes convolution, group normalization (GN)\cite{Wu2018} and Leaky ReLU activation, to progressively extract features. Suppose the number of output channels is $C$, then we set the first convolution layer outputs $\frac{C}{2}$ channels, and the second and third convolution layers output $\frac{C}{4}$ channels. All these outputs are concatenated together, followed by $1 \times 1$ convolution layers for feature fusion. Here, we adopts two $1 \times1$ convolution layers to reduce the number of parameters (Params.) and increase the nonlinearity. The first one compresses the channels to $\frac{C}{4}$ and the second one expands the channels back to $C$, which constitutes a bottleneck structure. Then, the fused features are fed to another $3 \times 3$ convolution layer with sigmoid activation to determine the attention map, which is multiplied with the fused features to highlight more important ones and suppress less useful ones. Finally, the calibrated features are added to the input to form the residual connection. 

This simple structure integrates both multi-scale feature fusion and attention mechanism. The outputs of the first three convolution layers have different receptive fields, which is equivalent to use convolutions with different kernel sizes to extract features. The concatenation and $1\times 1$ convolutions aims to fuse the features with different scales. Then, spatial attention is adopted to adaptively determine the informative features for better feature representation, with only a little extra computation cost. Our MARB is very lightweight compared with the common residual block (RB) with two $3\times 3$ convolution layers, each of which outputs $C$ channels. The Params. of the two types of blocks are calculated as follows
\begin{align*}
P_\text{MARB}&=C\times 3\times 3\times \frac{C}{2}+\frac{C}{2}\times 3\times 3\times \frac{C}{4}+\frac{C}{4}\times 3\times 3\times \frac{C}{4}\\&+C\times 1\times 1\times \frac{C}{4}\times 2+C\times 3\times 3\times 1\\&=\frac{107}{16}C^2+9C,\\
P_\text{RB}&=C\times 3\times 3\times C\times 2=18C^2.
\end{align*}
Then, we can obtain the ratio $r=\frac{P_\text{MARB}}{P_\text{RB}}=\frac{6.69C+9}{18C}$. As long as $C>0.796$, $r<1$. The channel numbers we set are between $16$ and $256$. Therefore, MARB has much fewer parameters than RB given the same input and output channel numbers.

\subsection{Multi-level pyramid denoising network}

\begin{figure*}[ht]
\centering
\includegraphics[width=1\textwidth]{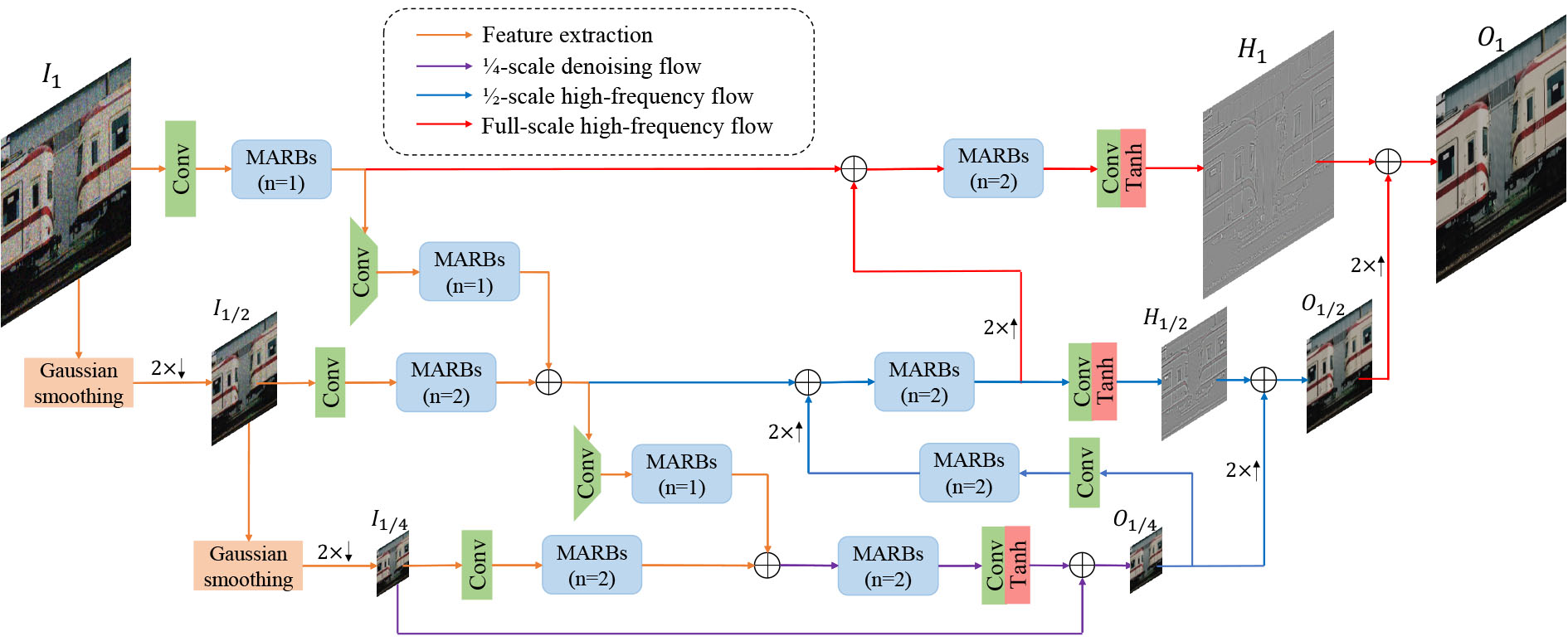} 
\caption{The architecture of MPDNet, which consists of a $\frac{1}{4}$-scale denoising flow (purple lines), a $\frac{1}{2}$-scale high-frequency flow (blue lines) and a full-scale high-frequency flow (red lines). These three flows share the same feature extraction process (orange lines).}
\label{fig:MPDNet}
\end{figure*}

The raw images with low photon counts are usually corrupted by severe noise. It is not easy to learn a direct mapping from noisy images to clean images. To reduce the learning difficulty, we develop the MPDNet, which adopts multi-level learning approach to estimate the small-scale noise map and larger-scale high-frequency maps. We draw on the idea of Laplace pyramid, which can decompose an image into a low-frequency component at the top level and a series of high-frequency components with different scales. Specifically, the $l$th high-frequency component $h_l\in \mathbb {R}^{\frac{h}{2^l}\times \frac{w}{2^l}}$ is obtained by $h_l=t_l-up_g(down_g(t_l))$, where $t_l$ is the image at the $l$th level, $up_g(\cdot)$ and $down_g(\cdot)$ denote $2\times$ up-sampling and down-sampling with Gaussian smoothing, and the original image $t_0$ has a shape of $h\times w$. With Laplace pyramid, the original image can be reconstructed without distortion.

Our MPDNet shown in Fig.~\ref{fig:MPDNet} consists of three levels, or three information flows, including the $\frac{1}{4}$-scale denoising flow (purple lines), $\frac{1}{2}$-scale high-frequency flow (blue lines) and full-scale high-frequency flow (red lines). These three flows share the same feature extraction process (orange lines). 

The feature extractions are conducted on the multi-scale input images, i.e. $I_1$, $I_{\frac{1}{2}}$, $I_{\frac{1}{4}}$, to encode the contextual information with different scales. $I_\frac{1}{2}$ and $I_\frac{1}{4}$ are obtained by down-sampling the input image $I_1$. Gaussian smoothing is used before each down-sampling to avoid the aliasing effect. $I_1$ is first fed to a convolution layer and an MARB to obtain the full-scale features $F_1$. $F_1$ further passes through a convolution layer with $\text{stride}=2$ and an MARB, and the output is added with the features extracted from $I_\frac{1}{2}$ by a convolution layer and two MARBs, to obtain the $\frac{1}{2}$-scale features $F_\frac{1}{2}$. Similarly, $F_\frac{1}{2}$ is further down-sampled and then added with the features extracted from $I_\frac{1}{4}$ to obtain the $\frac{1}{4}$-scale features $F_\frac{1}{4}$. In this way, the features with different scales and levels are aggregated to better guide the subsequent learning of noise suppression and high-frequency details. 

In the denoising flow, $F_\frac{1}{4}$ passes through two MARBs and a convolution layer with Tanh activation to estimate the noise map, which is added with $I_\frac{1}{4}$ to yield the $\frac{1}{4}$-scale output image $O_\frac{1}{4}$. Usually, clean images can provide better features for the learning of high-frequency components. Thus, in the $\frac{1}{2}$-scale high-frequency flow, the features of the denoised image $O_\frac{1}{4}$ are first extracted, and then up-sampled by twice and added with $F_\frac{1}{2}$ to supplement the high-resolution features. Next, these features are fed to the following two MARBs and a convolution layer with Tanh activation to obtain the $\frac{1}{2}$-scale high-frequency map $H_\frac{1}{2}$. As with the Laplace pyramid, $\frac{1}{2}$-scale output image $O_\frac{1}{2}$ is obtained by up-sampling $O_\frac{1}{4}$ and adding $H_\frac{1}{2}$. In the full-scale high-frequency flow, the features before the last convolution layer in the $\frac{1}{2}$-scale high-frequency flow are up-sampled and then added with $F_1$, followed by the subsequent MARBs and convolution layer to yield the full-scale high-frequency map $H_1$. The final output image $O_1$ is obtained by up-sampling $O_\frac{1}{2}$ and adding $H_1$. 

The information flows of MPDNet can be represented as
\begin{align*}
I_{\frac{1}{2}}&=down(g(I_1)), \ I_{\frac{1}{4}}=down(g(I_{\frac{1}{2}})),\\ 
O_{\frac{1}{4}}&=f_\text{D}(I_1,I_{\frac{1}{2}},I_{\frac{1}{4}})+I_{\frac{1}{4}},\\
H_{\frac{1}{2}}&=f_{\frac{1}{2}\text{HF}}(I_1,I_{\frac{1}{2}},I_{\frac{1}{4}}), \ H_1=f_\text{HF}(I_1,I_{\frac{1}{2}},I_{\frac{1}{4}}),\\
O_{\frac{1}{2}}&=up(O_{\frac{1}{4}})+k\times H_{\frac{1}{2}}, \ O_1=up(O_{\frac{1}{2}})+k\times H_1,
\end{align*}
where $g(\cdot)$ means Gaussian smoothing, $up(\cdot)$ and $down(\cdot)$ mean $2\times$ up-sampling and down-sampling operations, $f_\text{D}$ denotes the denoising flow, $f_{\frac{1}{2}\text{HF}}$ and $f_\text{HF}$ denote $\frac{1}{2}$-scale and full-scale high-frequency flows, respectively, and $k$ is a factor that can be manually designated during testing to control the image sharpness.

Our MPDNet can ease the learning of severe noise suppression. Since it is much easier to obtain small-scale clean and smooth images, denoising is only conducted on the $\frac{1}{4}$-scale images. However, the up-sampled small-scale images are always blurring. Thus, high-frequency components need to be learned to supplement the details. The high-frequency details are mainly reflected on the object boundaries and textures, and learning these sparse structures are definitely easier than learning the completed image features. Our MPDNet can be trained end-to-end to learn the noise removal and high-frequency details simultaneously.

\subsection{Luminance adjustment module}

\begin{figure*}[ht]
\centering
\includegraphics[width=1\textwidth]{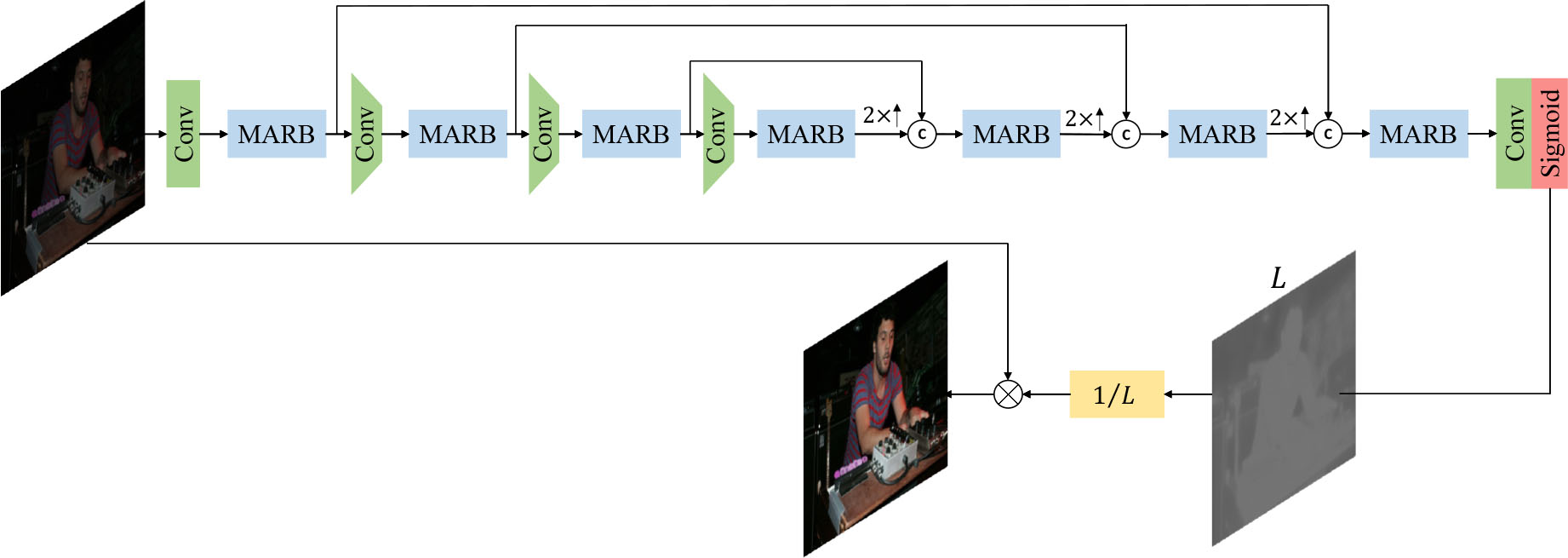} 
\caption{The architecture of LA module, which estimates the illumination of the input image. The inverse of the illumination is multiplied with the input image to obtain the normally exposed image.}
\label{fig:LA_full}
\end{figure*}

\begin{figure}[ht]
\centering
\includegraphics[width=0.45\textwidth]{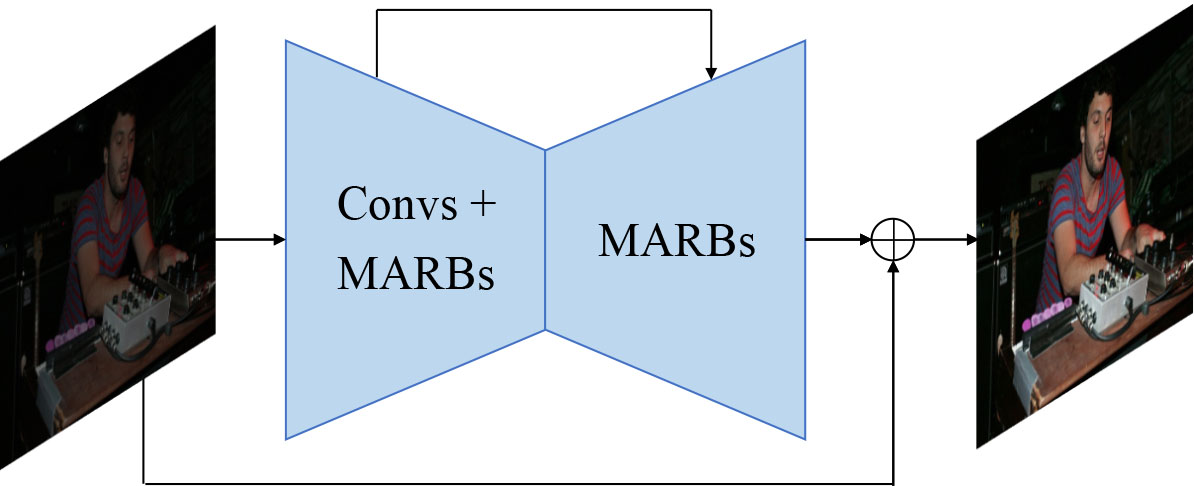} 
\caption{Another architecture for luminance adjustment, which adopts residual learning approach.}
\label{fig:LA_residual}
\end{figure}

Pool lighting conditions can seriously influence the image quality. In our framework, the images have been denoised by the MPDNet, so we only need to consider luminance enhancement here. To reduce the color distortion, our LA module learns to estimate the illumination of the input image. Since the pixels with lower illumination should be magnified by larger values and the pixels with higher illumination should be magnified by smaller values, the input image can directly multiply with the inverse of the illumination to obtain the normally exposed image. As the illumination has no ground truth, it is learned by a unsupervised approach with normal-light image as the implicit supervisory signal. 

As shown in Fig.~\ref{fig:LA_full}, our LA module adopts a simple encode-decoder architecture, with convolution layers ($\text{stride}=2$) to down-sample the feature maps, MARBs to extract features at different scales and skip connections to fuse high-resolution and high-level semantic features. The input of LA module is the denoised image from the MPDNet and the output is the estimated illumination $L$, each element value of which is between $0$ and $1$. The illumination is expected to be globally smooth while preserving the object structures. Then, the inverse of illumination $L^{-1}$ is multiplied with the input to yield the final enhanced image. In this way, the output image has little color distortion as the color information of the input image is fully reserved. The final restored image $R$ by our framework is obtained by
\begin{align*}
R=O_1\times L^{-1}.
\end{align*}

In addition, we consider another architecture for luminance adjustment, as shown in Fig.~\ref{fig:LA_residual}. It adopts the same encoder-decoder architecture as our LA module, but predicts the residual image rather than the illumination. The performance comparison will be illustrated in sub-section~\ref{sub_sec:ablation}.

\subsection{Loss functions}
The loss functions for the MPDNet include the $\ell_1$ loss of high-frequency components
\begin{align}
\ell_H=\sum_{s=1,\frac{1}{2}}\left\|H_s-\hat{H}_s\right\|_1,
\end{align}
and the $\ell_1$ loss of output images with different scales 
\begin{align}
\ell_O=\sum_{s=1,\frac{1}{2},\frac{1}{4}}\left\|O_s-\hat{O_s}\right\|_1,
\end{align}
where $\hat{H}_\frac{1}{2}$, $\hat{H}_1$, $\hat{O}_\frac{1}{4}$ and $\hat{O}_\frac{1}{2}$ are the corresponding ground truths obtained by the Laplace pyramid.

In addition, we introduce SSIM loss\cite{Wang2004} $\ell_\text{SSIM}$ and perceptual loss\cite{Ledig2017} $\ell_\text{vgg}$ for $O_1$, similar with\cite{Zhang2021a}, to further improve the visual quality of the output images. Then, the full loss to train the MPDNet is formulated as
\begin{align}\label{eq:MPD loss}
\ell_\text{MPD}=\ell_H+\ell_O+\ell_\text{SSIM}(O_1,\hat{O}_1)+\ell_\text{vgg}(O_1,\hat{O}_1).
\end{align}

The loss functions for the LA module comprises of the $\ell_1$ loss between the enhanced image $R$ and its ground truth $\hat{R}$, and the structure-preserved smooth loss for the illumination $L$, which is defined as 
\begin{align}
\ell_\text{smooth}(L)=|\nabla L|\times \exp(-\eta |\nabla \hat{R}|),
\end{align}
where $\nabla$ denotes the gradient along both the horizontal and vertical directions, and $\eta$ is a hyperparameter to adjust the degree of structure preservation. The structure reference is provided by the ground truth $\hat{R}$, so that the illumination could preserve the structure boundaries of the ground truth image. Thus, the full loss to train the LA module is formulated as
\begin{align}\label{eq:LA loss}
\ell_\text{LA}=\ell_1(R,\hat{R})+\ell_\text{smooth}(L).
\end{align}

\section{Experiments}\label{sec:experiments}

\subsection{Simulation of low-photon-count imaging}

\begin{figure*}
\centering
\includegraphics[width=1\textwidth]{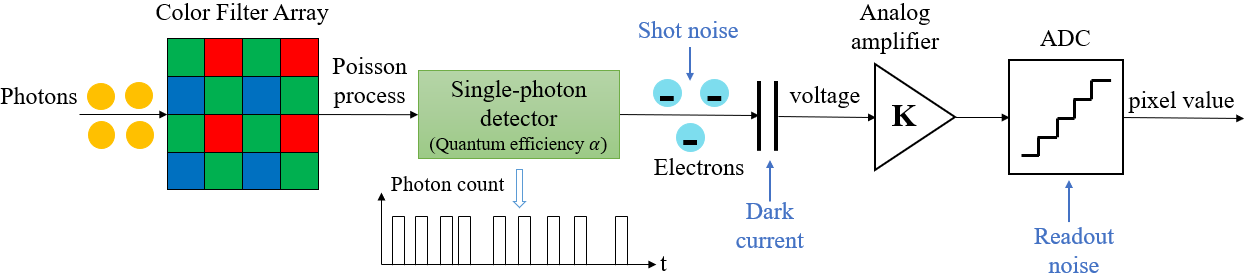} 
\caption{Single-photon imaging process. The photons arrive at the color filter array for wavelength selection and then the single-photon detector to be converted to the electrons. The electrical signal is further processed by the analog amplifier and ADC to yield the pixel values. The photon arrival follows Poisson process, which leads to the shot noise. Dark current and readout noise are also introduced during the other stages of imaging.}
\label{fig:imaging}
\end{figure*}

\begin{figure*}
\centering
\includegraphics[width=1\textwidth]{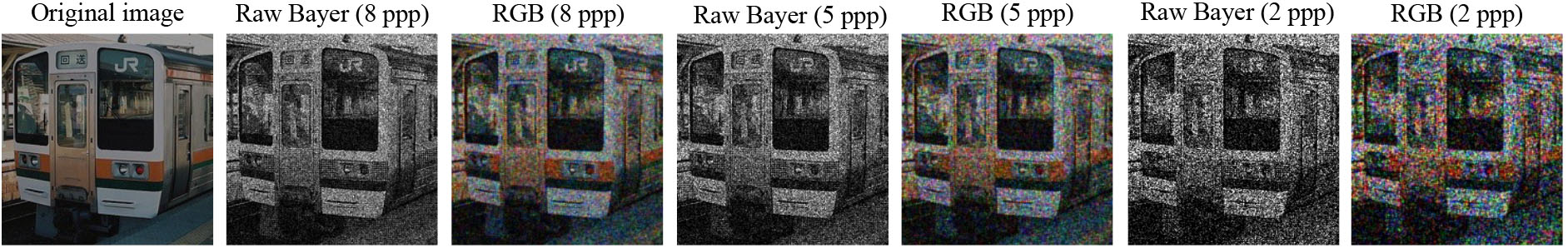} 
\caption{An original image and its corresponding synthetic raw Bayer images and RGB images with different average photon counts.}
\label{fig:example}
\end{figure*}

We adopt a similar method with\cite{Elgendy2021} to simulate the low-photon-count imaging. The imaging process is described in Fig.~\ref{fig:imaging}. When the photons arrive at the camera, the color filter array, usually with Bayer pattern, only allows the photons with corresponding color to pass through. Then, the selected photons reach the single-photon detector, which has the photon-counting capability, and are converted to the electrons. We assume the single-photon detector has a quantum efficiency $\alpha$, which represents the percentage of photons detected by the sensor. The electrical signal is further processed by the analog amplifier and analog-digital conversion (ADC) to yield the pixel values. This imaging process introduces different types of noises during different stages, including the shot noise, dark current and readout noise. The dominant noise in the low-photon-count case is the shot noise, caused by the Poisson process of photon arrival from scene to sensor.

The simulation for low-photon-count imaging should operate on the Bayer pattern of each original image, and we formulate it in the following equation,
\begin{align}\label{eq:image synthesis}
X_\text{syn}=K\Big(\underbrace{\text{Poisson}\big(X_\text{Bayer}/\text{mean}(X_\text{Bayer})\times p \big)}_{\text{photon\ count}}+n_r\Big),
\end{align}
where $X_\text{Bayer}$ is the original Bayer image, $p$ is a factor to control the photon level, and $n_r$ is the readout noise obeying Gaussian distribution. Considering the low readout noise and dark current of real QIS, the standard deviation of readout noise is set to $0.25e^{-}$ and the dark current is ignored. $K$ means scaling the pixel values to $0\sim255$ to obtain the raw Bayer image $X_\text{syn}$. Then, we use the interpolation demosaicing operation to obtain the corresponding RGB images, which are then input to our framework for restoration.

We selected $8000$ images from Pascal VOC2012 dataset for training and $500$ images for validation. The original images are treated as the ground truths and the corresponding degraded images are synthesized by Eq.~\ref{eq:image synthesis}. The average photon count for each image is randomly sampled from the range $[1,10]$ during training and validation. Our testing data contain $180$ images from Pascal VOC2012 dataset, $150$ images from COCO val2017 dataset and $24$ images from Kodak dataset, with photon counts ranging from $1$ to $10$. We present several examples of synthetic raw Bayer images and RGB images with different average photon counts corresponding to one original image, as shown in Fig.~\ref{fig:example}. It can be seen that the noise level becomes higher with the decrease of average photon count due to the lower signal-to-noise ratio.

\subsection{Implementation details}
We first trained the MPDNet with channel number for feature extraction expanding from $32$ to $256$. During training, the input images were synthesized from the ground truth images by random average photon counts between $1$ and $10$, and cropped to $256\times 256$ patches randomly. Moreover, they are augmented with random horizontal flips. The Adam optimizer with a initial learning rate of $1\times 10^{-3}$ and a decay rate of $0.8$ every $30$ epochs was used. The MPDNet was trained for $150$ epochs with batch size of $16$.

Next, we trained the LA module with channel number increasing from $16$ to $128$ in the encoder. During training, the input underexposed images were synthesized from the ground truth images by adjusting the $V$ channel of the HSV color space with a random factor between $0.1$ to $0.9$. The random cropping and flips were still adopted for data augmentation. We set the initial learning rate to $1\times 10^{-3}$ with a decay rate of $0.8$ every $20$ epochs. The LA module was trained for $80$ epochs with batch size of $16$. Our framework was implemented by PyTorch on Nvidia Tesla V100-SXM2 GPU.

\subsection{Ablation study}\label{sub_sec:ablation}
In this subsection, we perform ablation study on our MPDNet and LA module.

\begin{figure*}[ht]
\centering
\includegraphics[width=1\textwidth]{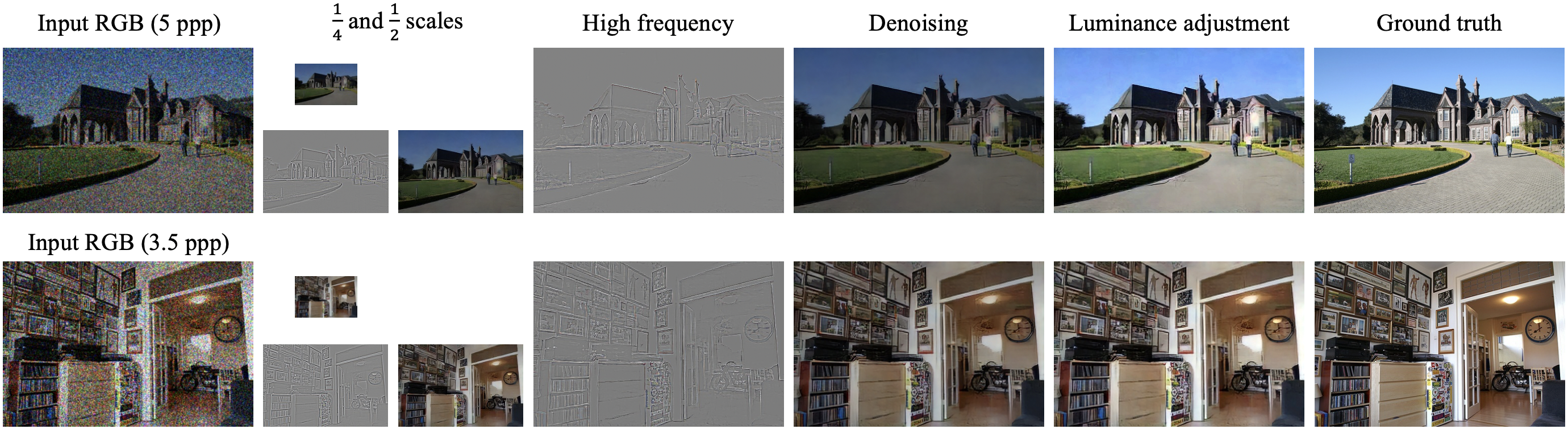} 
\caption{Two restoration results of our framework, including the intermediate outputs of the MPDNet, i.e., the $\frac{1}{4}$-scale denoised images, $\frac{1}{2}$-scale high-frequency images, $\frac{1}{2}$-scale denoised images and full-scale high-frequency images.}
\label{fig:full outputs}
\end{figure*}

We first provide two restoration results of our framework, as shown in Fig.~\ref{fig:full outputs}. The input images are with $5$ ppp and $3.5$ ppp, respectively, which are with low photon counts. We also present the intermediate outputs of our MPDNet, including the $\frac{1}{4}$-scale denoised images, $\frac{1}{2}$-scale high-frequency images, $\frac{1}{2}$-scale denoised images and full-scale high-frequency images. As can be seen, our MPDNet can output smooth $\frac{1}{4}$-scale images and effectively predict the $\frac{1}{2}$-scale and full-scale high-frequency components to compose the high-quality denoised images. As the denoised images may be with poor luminance, they are further fed to the LA module for luminance enhancement. Thus, we can obtain more visually compelling images, which have similar luminance and little color distortion compared with their ground truths. It should be noted that our LA module has enough flexibility since it adjusts more under poor lighting conditions and only adjusts little under relatively good lighting conditions, as presented in these two examples.

\subsubsection{MPDNet}

\begin{table}[t]
\caption{Ablation study on the MPDNet structure}
\centering
\begin{tabular}{p{3.6cm}|cc|cc}
\toprule
Configurations  &Params. & FLOPs & PSNR & SSIM\\[0.5ex]
\midrule
Our MPDNet  & \SI{1.709}{M} & \SI{15.431}{G}  &\textbf{28.36} & \textbf{0.8270} \\
Replace MARB with RB  & \SI{2.446}{M} & \SI{21.880}{G}  &28.23 & 0.8257 \\
w/o multi-scale input images  & \SI{1.212}{M} & \SI{12.045}{G} & 27.47 & 0.8081\\
w/o multi-level structure & \SI{1.543}{M} & \SI{14.543}{G} & 27.26 & 0.7918\\
\bottomrule
\end{tabular}
\label{tab:MPDNet structure ablation}
\end{table}

\begin{figure*}[ht]
\centering
\includegraphics[width=1\textwidth]{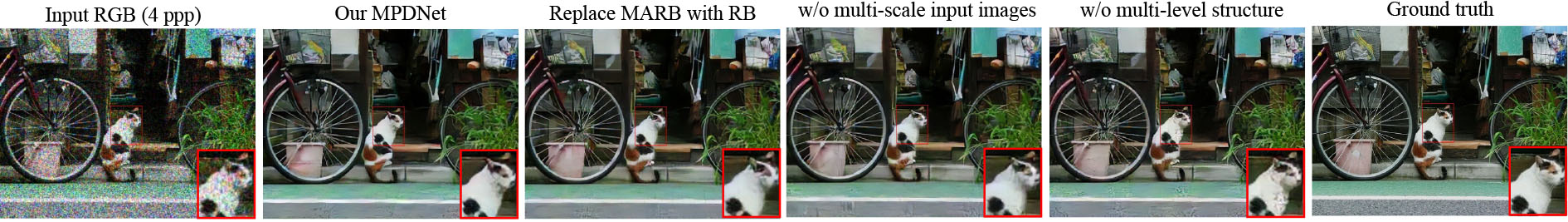} 
\caption{Visual results of ablation study on the MPDNet structure, including the input image with $4$ ppp, denoising results of different configurations and ground truth.}
\label{fig:MPDNet structure ablation}
\end{figure*}

\begin{table}[t]
\caption{Ablation study on the MPDNet loss terms}
\centering
\begin{tabular}{p{3.2cm}|cc}
\toprule
Configurations  & PSNR & SSIM\\[0.5ex]
\midrule
Full loss                &\textbf{28.36} & \textbf{0.8270} \\
w/o high-frequency loss  & 27.44 & 0.8053 \\
w/o SSIM loss            & 28.03 & 0.8010 \\
w/o perceptual loss      & 28.18 & 0.8217 \\
\bottomrule
\end{tabular}
\label{tab:MPDNet loss ablation}
\end{table}

\begin{figure*}[ht]
\centering
\includegraphics[width=1\textwidth]{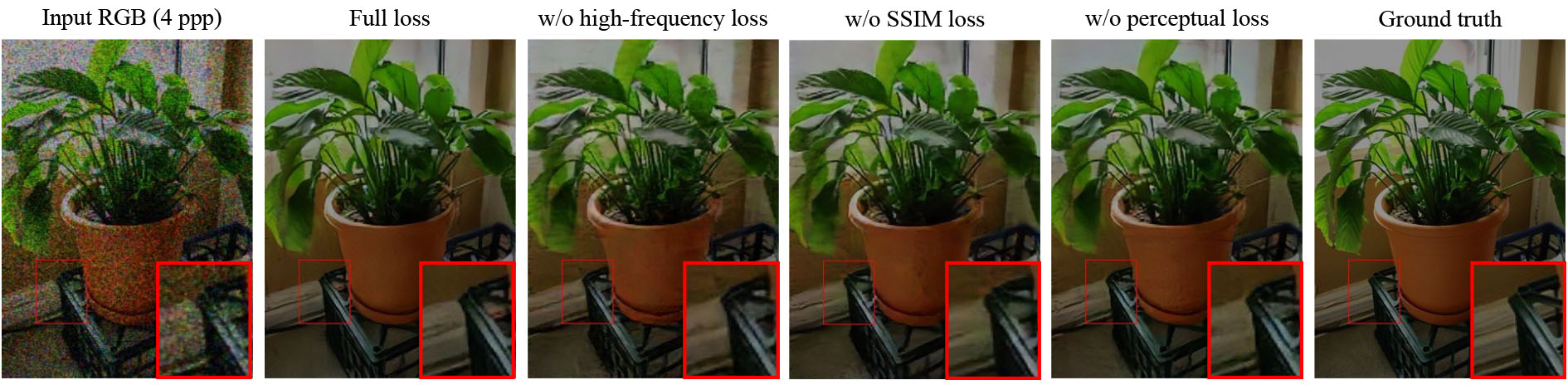} 
\caption{Visual results of ablation study on the loss terms for training MPDNet, including the input image with $4$ ppp, denoising results of different configurations and ground truth.}
\label{fig:MPDNet loss ablation}
\end{figure*}

Firstly, we conduct an ablation study on our MPDNet by comparing the performance of several different architectures. We use PSNR and SSIM as the quantitative metrics, which are calculated on the test sets containing the raw images with uniformly distributed average photon counts from $1$ to $10$. We first test our MPDNet and its results are treated as the comparison object. The quantitative results as well as the Params. and FLOPs of different configurations are recorded in Table~\ref{tab:MPDNet structure ablation}. 

To study the advantage of our MARB, we replace all the MARBs with RBs, i.e., the common residual blocks with the intermediate channel number being the half of output channel number to reduce the model size. As shown in Table~\ref{tab:MPDNet structure ablation}, this configuration (`Replace MARB with RB') achieves a little lower PSNR and SSIM than our MPDNet but with more parameters and computation, which reflects the effectiveness and efficiency of our MARB.

To investigate the benefit of feature extraction on multi-scale input images, we remove the corresponding convolution layers and MARBs after $I_{\frac{1}{2}}$ and $I_{\frac{1}{4}}$. Hence, only the features of full-scale image $I_1$ are extracted. It can be seen that the results of this configuration (`w/o multi-scale input images') are much lower than our MPDNet even if the Params. and FLOPs are reduced obviously. Therefore, the feature extraction on multi-scale images contributes to better denoising performance because of the encoding of richer contextual information. 

Our MPDNet adopts an multi-level learning strategy with three information flows to learn the noise map and high-frequency maps, respectively. Thus, we investigate the benefit of this multi-level structure by building and testing another network, which has similar architecture with our MPDNet in general but directly outputs the full-scale denoised images without the multi-level learning. It can be seen that this configuration (`w/o multi-level structure') has the worst results among all the others, which suggests that the direct noisy-to-clean mapping can not suppress noise effectively when the noise level is high. 

We present the denoising results of different configurations for one image with $4$ ppp, as shown in Fig.~\ref{fig:MPDNet structure ablation}. We can see that our MPDNet can achieve better denoising performance than the others, with less remaining noise and finer object details.

Then, we perform another ablation study on the loss terms for training MPDNet. In our MPDNet, we explicitly designate two information flows to learn the high-frequency components. What if we remove the high-frequency supervision? Thus, the network only gradually refines the outputs with increased resolution but does not specifically learn the high-frequency details. From Table~\ref{tab:MPDNet loss ablation}, we can see that the performance is obviously degraded when the high-frequency loss is not included. Therefore, explicit high-frequency supervision is beneficial for noise removal as it enforces the network to learn the clean high-frequency components.

Moreover, we check the effect of SSIM loss and perceptual loss. When the SSIM loss is removed, the SSIM has obvious decline. When the perceptual loss is not included, the results also decline slightly. Fig.~\ref{fig:MPDNet loss ablation} presents the visual results of the MPDNet trained by different loss settings. It can be seen that the full loss can achieve best denoising performance, while the MPDNet without high-frequency supervision can not suppress noise well in some regions so that the output image has some unexpected artifacts.

\subsubsection{LA module}

\begin{table}[t]
\caption{Ablation study on the LA module}
\centering
\begin{tabular}{p{3.1cm}|cc|cc}
\toprule
Configurations  &Params. & FLOPs & PSNR & SSIM\\[0.5ex]
\midrule
Our LA module      & \SI{0.3149}{M} & \SI{1.511}{G}  & \textbf{24.79} & \textbf{0.7779}\\
Residual learning approach & \SI{0.3152}{M} & \SI{1.530}{G} & 24.24 & 0.7715\\
\bottomrule
\end{tabular}
\label{tab:LA ablation}
\end{table}

\begin{figure*}[ht]
\centering
\includegraphics[width=1\textwidth]{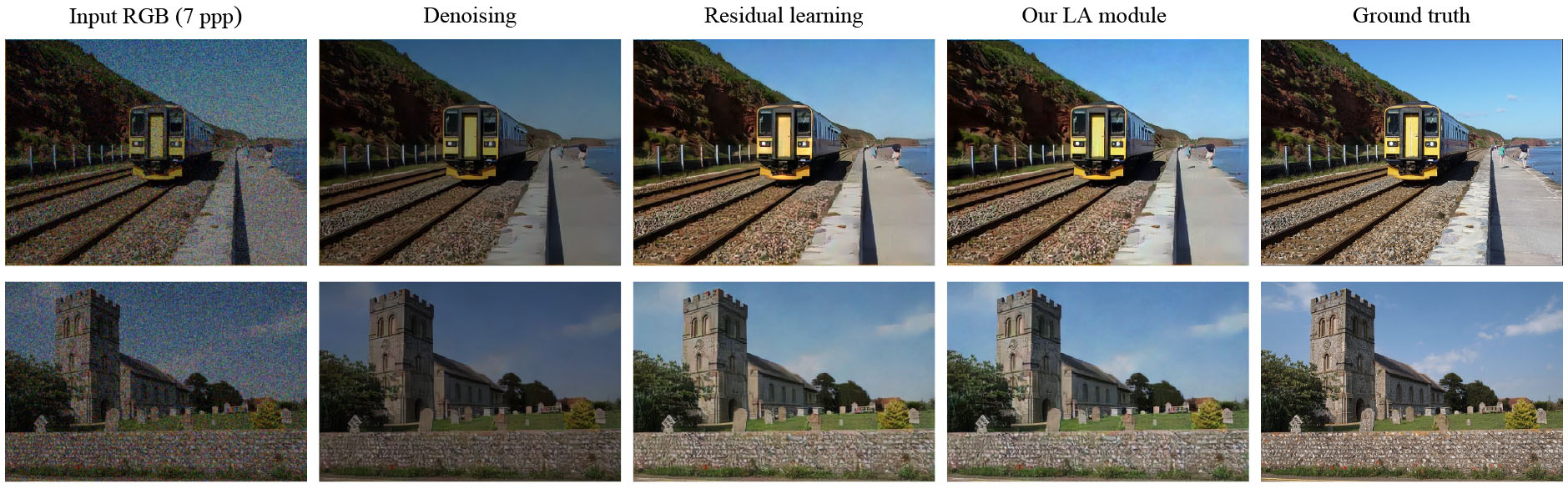} 
\caption{Visual comparisons of luminance adjustment, including the input images with $7$ ppp, denoising outputs, luminance adjustment outputs and ground truths.}
\label{fig:LA ablation}
\end{figure*}

Here, we compare the performance of our LA module and the residual learning approach in Fig.~\ref{fig:LA_residual}. They follow the MPDNet to further enhance the denoised images. Table~\ref{tab:LA ablation} shows that our LA module obtain better results, and also has a little fewer Params. and FLOPs as it only predicts the illumination map with one channel rather than the residual image with three channels. 

We provide two visual comparisons, as shown in Fig.~\ref{fig:LA ablation}, including the input images with $7$ ppp, denoising outputs, luminance adjustment outputs of the two architectures and ground truths. We can see that the results of residual learning approach has more color distortion (e.g. the train in the first image and the sky in the second image) than our LA module. Therefore, for the clean images, luminance adjustment through illumination estimation is more effective than predicting the residual images. In short, our LA module is very lightweight and introduces little extra computation burden to the framework.

\subsection{Comparisons}

\begin{figure*}[ht]
\centering
\includegraphics[width=1\textwidth]{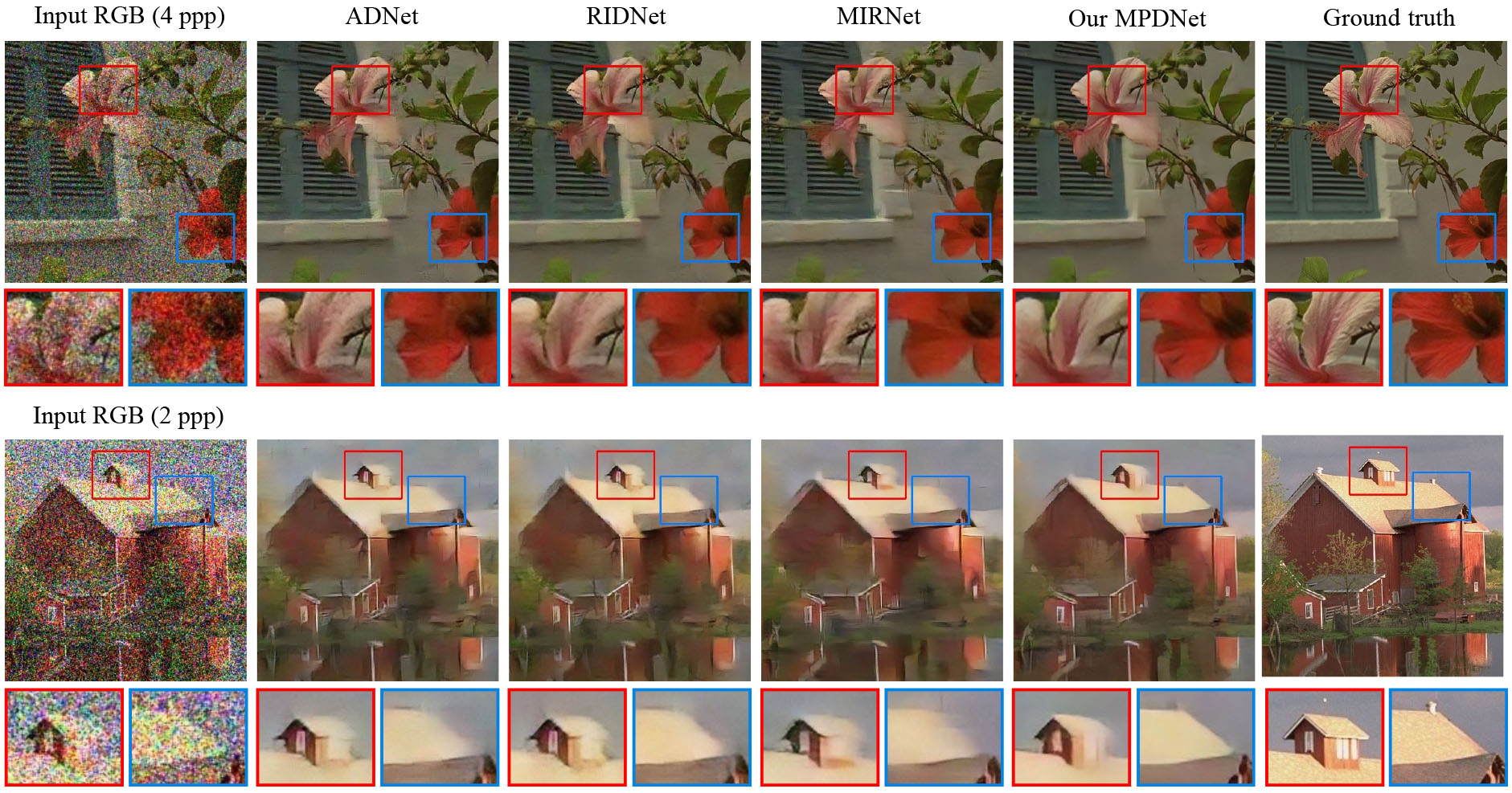} 
\caption{Visual comparison of denoising performance among different methods.}
\label{fig:denoising comparison}
\end{figure*}

\subsubsection{Comparison with other denoising methods}

\begin{table}[t]
\caption{Quantitative results of different denoising methods}
\centering
\begin{tabular}{c|cc|ccc}
\toprule
Methods  &Params. & FLOPs & PSNR & SSIM & LPIPS$\downarrow$\\[0.5ex]
\midrule
Our MPDNet             & \SI{1.709}{M} & \SI{15.431}{G}  & \textbf{28.36}  & \textbf{0.8270} &\textbf{0.1726}\\
ADNet\cite{Tian2020}   & \SI{1.169}{M} & \SI{76.631}{G}  & 26.98 & 0.7765 & 0.2237\\
RIDNet\cite{Anwar2019} & \SI{1.499}{M} & \SI{98.126}{G}  & 27.24 & 0.7966 & 0.2081\\ 
MIRNet\cite{Zamir2020} & \SI{2.660}{M} & \SI{66.565}{G}  & 27.95 & 0.8089 & 0.1876\\
\bottomrule
\end{tabular}
\label{tab:denoising comparison}
\end{table}

\begin{figure*}[ht]
\centering
\subfloat[PSNR]{\includegraphics[height=0.21\linewidth]{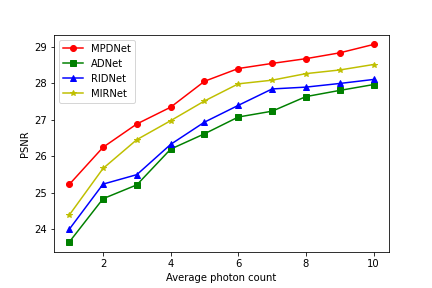}}
\hfil
\subfloat[SSIM]{\includegraphics[height=0.21\linewidth]{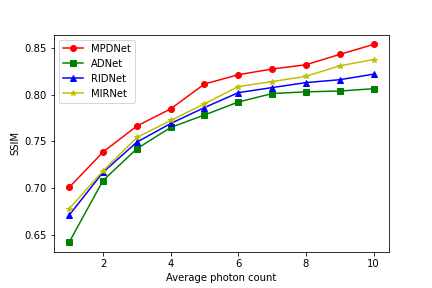}}
\hfil
\subfloat[LPIPS]{\includegraphics[height=0.21\linewidth]{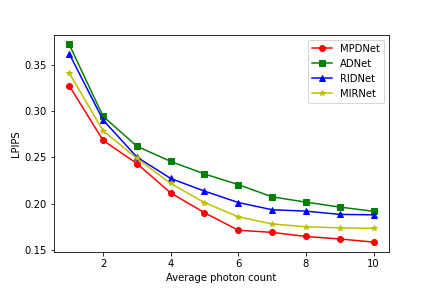}}
\caption{PSNR, SSIM and LPIPS of different denoising methods at various photon levels.}
\label{fig:plot}
\end{figure*}

Firstly, we compare our MPDNet with several other recently proposed blind denoising methods, including ADNet\cite{Tian2020}, RIDNet\cite{Anwar2019} and MIRNet\cite{Zamir2020}. These methods design different architectures to learn the full-scale residual map. We adjust the hyper-parameters of these networks to guarantee approximate Params. with our MPDNet. Then, we use the same datasets and training strategies to train them. The Params., FLOPs and quantitative results of these methods are recorded in Table~\ref{tab:denoising comparison}. We also calculate the learned perceptual image patch similarity (LPIPS) for more comprehensive evaluation. For this metric, smaller value is better. It can be seen that other denoising methods get worse quantitative results but with much more FLOPs compared with our MPDNet. Therefore, our MPDNet can achieve better denoising performance with higher efficiency. 

Table~\ref{tab:denoising comparison} gives the average results containing various photon levels. To compare the denoising performance at each photon level, we plot the PSNR, SSIM and LPIPS curves as a function of the average photon count ranging from $1$ to $10$, as shown in Fig.~\ref{fig:plot}. The values of all the metrics become better with the increase of average photon count while our MPDNet outperforms other methods obviously at any photon level. 

We present some visual comparisons among different denoising methods, as shown in Fig.~\ref{fig:denoising comparison}. The sizes of original images in Kodak dataset are large, so we only crop $384\times 384$ patches for display. The input images are with $4$ ppp and $2$ ppp, where some texture details are completely corrupted by the noise, especially for the case of $2$ ppp. The denoising visual comparison indicates that our MPDNet can restore more object details and edge information, as depicted by the zoom-in color-framed small patches, and get more smooth effect in the textureless regions, such as the wall and sky in the images. 

\subsubsection{Comparison of different frameworks for image restoration}

\begin{figure*}[ht]
\centering
\subfloat[Direct mapping approach]{\includegraphics[height=0.135\linewidth]{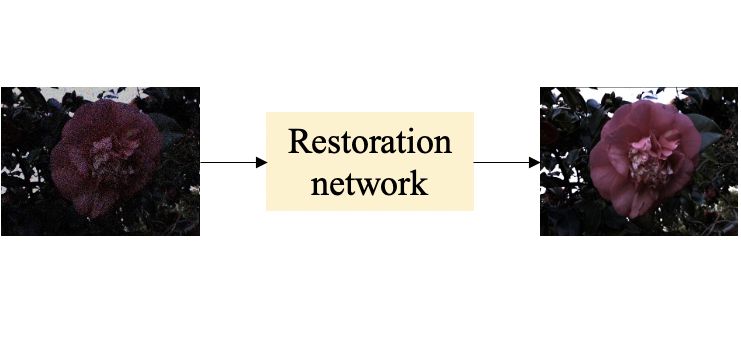}}
\hfil
\subfloat[Retinex-based approach]{\includegraphics[height=0.135\linewidth]{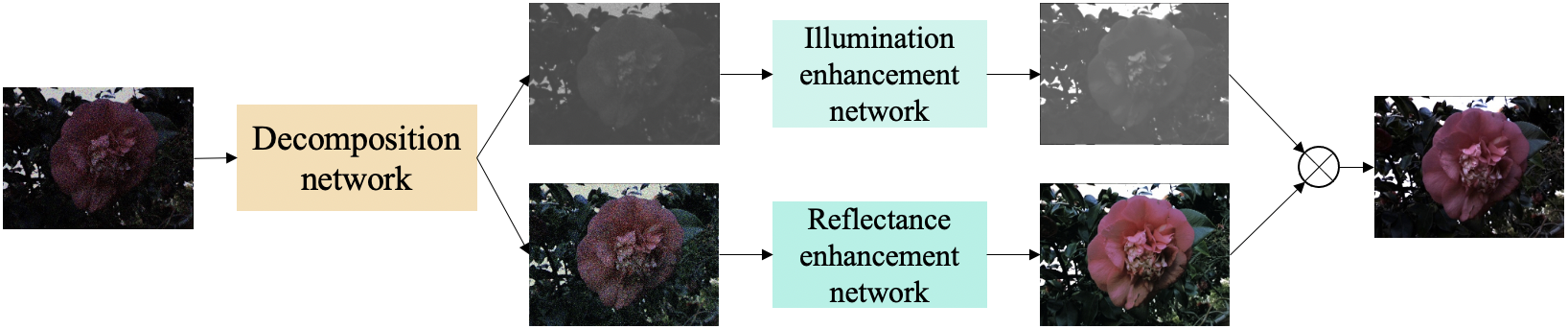}}
\caption{Other frameworks for low-light image enhancement. (a) Direct mapping from low-light noisy image to normal-light clean image. (b) Decomposition based on Retinex theory and enhancement for reflectance and illumination.}
\label{fig:other frameworks}
\end{figure*}

\begin{table}[t]
\caption{Quantitative results of different frameworks for image restoration}
\centering
\begin{tabular}{c|cc|ccc}
\toprule
Methods  &Params. & FLOPs & PSNR & SSIM &LPIPS$\downarrow$\\[0.5ex]
\midrule
Ours                 & \SI{2.024}{M} & \SI{16.942}{G}  & \textbf{24.79} & \textbf{0.7779} & \textbf{0.1994}\\
MBLLEN\cite{Lv2018}  & \SI{1.590}{M} & \SI{86.449}{G}  & 21.94 & 0.7345 & 0.2524\\
KinD\cite{Zhang2019} & \SI{2.197}{M} & \SI{19.857}{G}  & 21.85 & 0.7234 & 0.2739\\
\bottomrule
\end{tabular}
\label{tab:framework comparison}
\end{table}

\begin{figure*}[ht]
\centering
\includegraphics[width=1\textwidth]{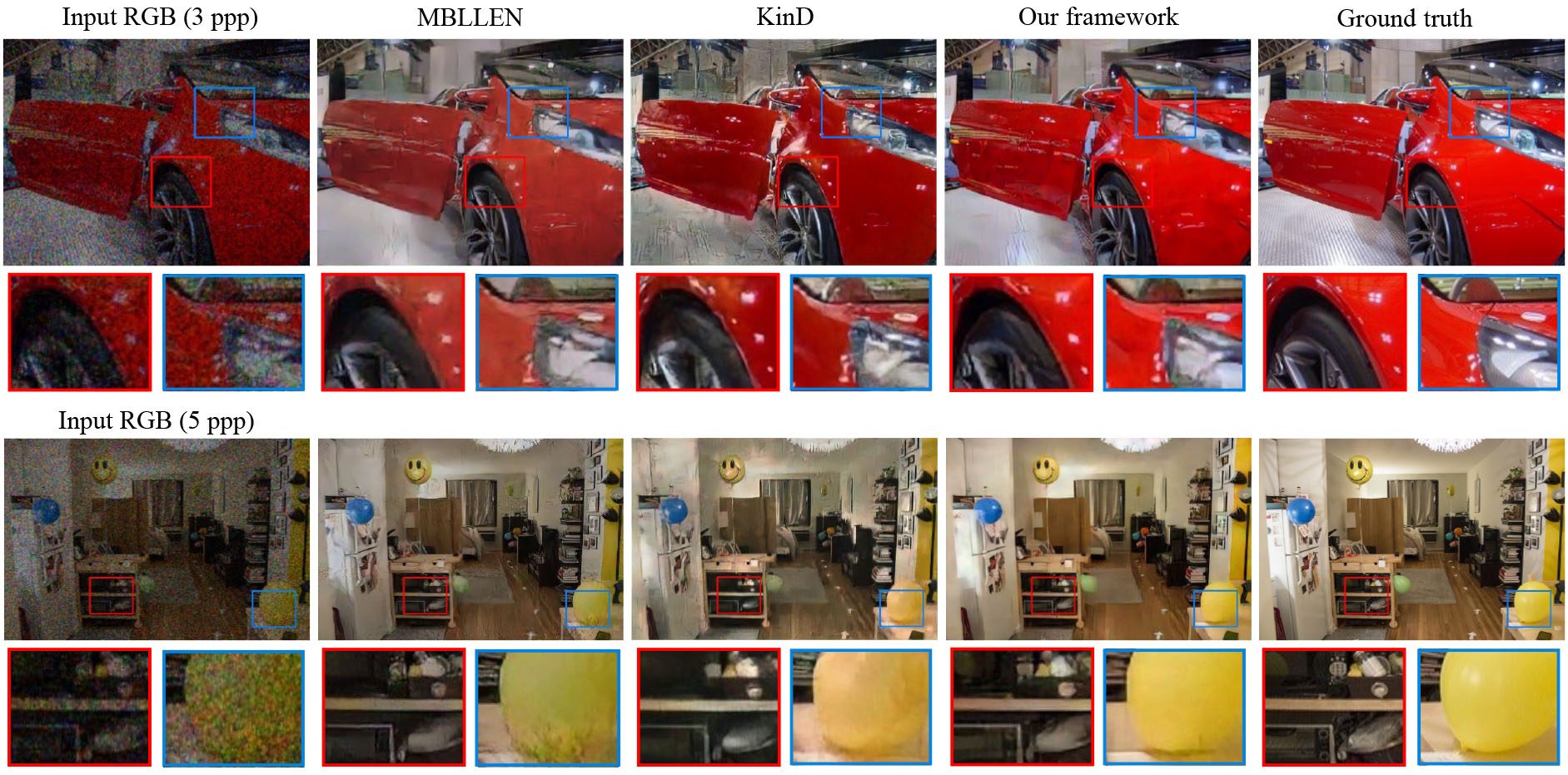} 
\caption{Visual comparison of different frameworks for image restoration.}
\label{fig:framework comparison}
\end{figure*}

As our framework can achieve both denoising and luminance enhancement, we make comparison with another two low-light image enhancement methods, MBLLEN\cite{Lv2018} and KinD\cite{Zhang2019}. MBLLEN adopts the direct mapping from low-light noisy image to normal-light clean image. KinD utilizes Retinex theory for decomposition and then enhances the reflectance and illumination respectively. These two approaches are illustrated in Fig.~\ref{fig:other frameworks}. 

Table~\ref{tab:framework comparison} indicates that our framework obtains the best quantitative results with the minimum FLOPs. We also present some visual comparisons among different frameworks for restoring images with low photon counts, as shown in Fig.~\ref{fig:framework comparison}. It can be seen that our framework can achieve better denoising and luminance recovery and therefore obtain images with better visual quality. By contrast, the MBLLEN outputs have some color distortion, with some regions unsmooth and still noisy. The KinD outputs also have obvious noise and color distortion. In short, our framework can achieve superior performance on image restoration with lightweight architecture and high efficiency.

\subsection{Extended experiment on the real degraded images}

\begin{figure}[ht]
\centering
\includegraphics[width=0.5\textwidth]{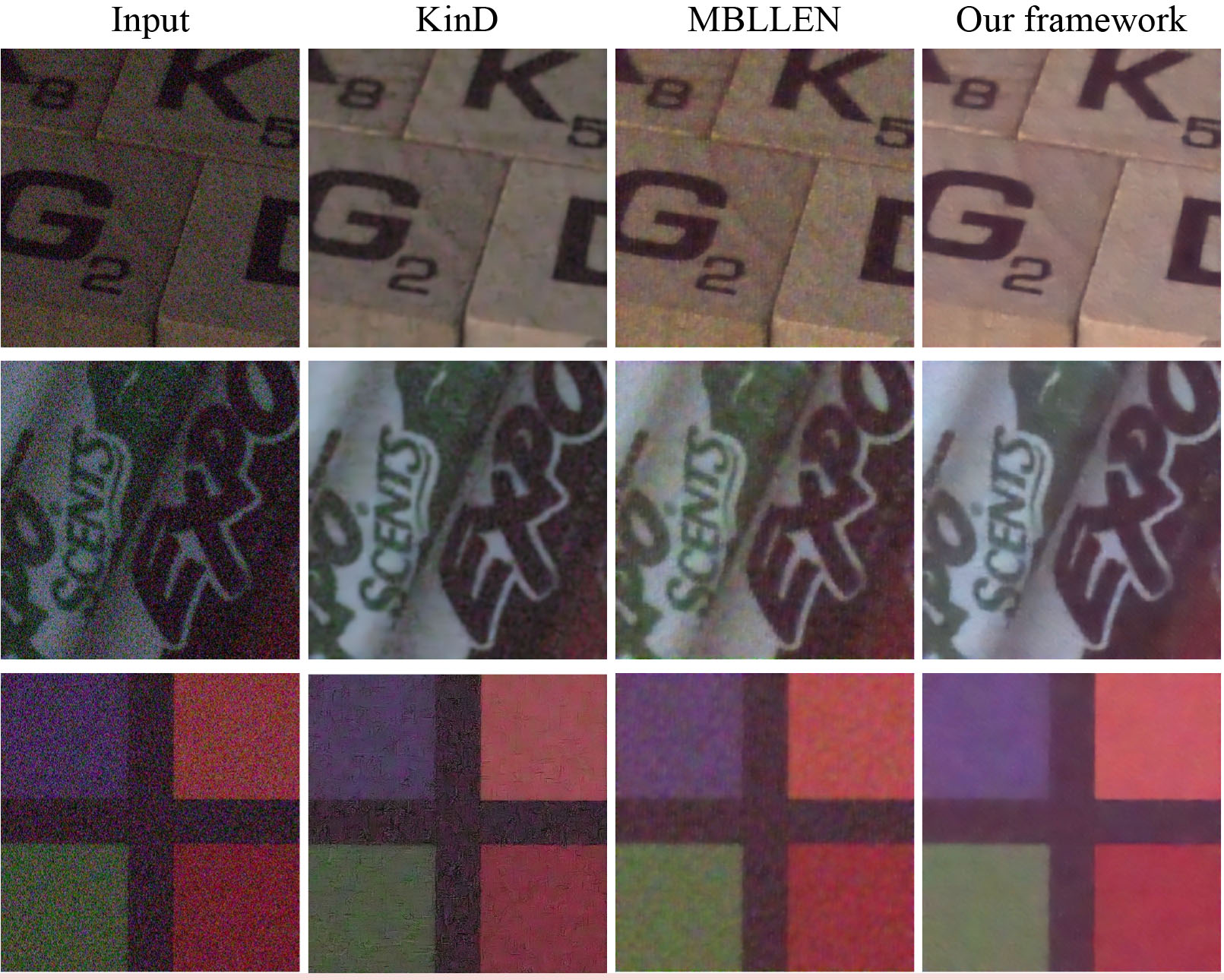} 
\caption{Visual comparison on the real degraded images.}
\label{fig:SIDD comparison}
\end{figure}

As we mentioned before, our framework is a generic solution for restoring images with noise and poor luminance. We also test on the real degraded images provided by\cite{Abdelhamed2018} to further verify its effectiveness. As the original images are with large sizes, we crop $512\times 512$ patches for testing to better show the restoration details. The visual comparison of different methods are presented in Fig.~\ref{fig:SIDD comparison}, which again suggests that the images restored by our framework have better visual quality with less noise and more proper luminance compared with the other methods, demonstrating the effectiveness and robustness of our framework for real image restoration.  

\section{Conclusion}\label{sec:conclusion}
In this paper, we study on the raw image restoration under low-photon-count imaging. The raw images with low photon counts are corrupted by severe noise and poor luminance. To reduce the recovery difficulty, we develop a framework to restore each image through separate denoising and luminance adjustment. Our framework consists of a MPDNet for noise suppression and an LA module for luminance adjustment. Our MPDNet employs a multi-level structure and utilizes the idea of Laplace pyramid to estimate the small-scale noise map and larger-scale high-frequency maps progressively. Feature extractions are performed on the multi-scale input images to encode richer contextual information. Our LA module aims to enhance the luminance of the image denoised by the MPDNet through estimating its illumination. The final restored image is obtained by multiplying the denoised image and the reverse of its estimated illumination. The main component of our MPDNet and LA module is the MARB, which integrates multi-scale feature fusion and spatial attention for better feature representation, with much fewer parameters than the common residual block. Extensive experiments have been conducted to verify the effectiveness and robustness of our framework for image restoration.    

\section*{Acknowledgment}
The work is supported in part by the Research Grants Council of Hong Kong (GRF 17201818, 17200019, 17201620) and the University of Hong Kong (104005864).

\bibliographystyle{IEEEtran}
\bibliography{references}

\begin{thebibliography}{10}
\providecommand{\url}[1]{#1}
\csname url@samestyle\endcsname
\providecommand{\newblock}{\relax}
\providecommand{\bibinfo}[2]{#2}
\providecommand{\BIBentrySTDinterwordspacing}{\spaceskip=0pt\relax}
\providecommand{\BIBentryALTinterwordstretchfactor}{4}
\providecommand{\BIBentryALTinterwordspacing}{\spaceskip=\fontdimen2\font plus
\BIBentryALTinterwordstretchfactor\fontdimen3\font minus
  \fontdimen4\font\relax}
\providecommand{\BIBforeignlanguage}[2]{{%
\expandafter\ifx\csname l@#1\endcsname\relax
\typeout{** WARNING: IEEEtran.bst: No hyphenation pattern has been}%
\typeout{** loaded for the language `#1'. Using the pattern for}%
\typeout{** the default language instead.}%
\else
\language=\csname l@#1\endcsname
\fi
#2}}
\providecommand{\BIBdecl}{\relax}
\BIBdecl

\bibitem{Elgendy2018}
O.~A. Elgendy and S.~H. Chan, ``Optimal threshold design for quanta image
  sensor,'' \emph{IEEE Transactions on Computational Imaging}, vol.~4, no.~1,
  pp. 99--111, 2018.

\bibitem{Ma2015}
J.~Ma and E.~R. Fossum, ``A pump-gate jot device with high conversion gain for
  a quanta image sensor,'' \emph{IEEE Journal of the Electron Devices Society},
  vol.~3, no.~2, pp. 73--77, 2015.

\bibitem{Chi2020}
Y.~Chi, A.~Gnanasambandam, V.~Koltun, and S.~H. Chan, ``Dynamic low-light
  imaging with quanta image sensors,'' in \emph{European Conference on Computer
  Vision (ECCV)}, 2020, pp. 122--138.

\bibitem{Abhiram2019}
A.~Gnanasambandam, O.~Elgendy, J.~Ma, and S.~H. Chan, ``Megapixel
  photon-counting color imaging using quanta image sensor,'' \emph{Optics
  Express}, vol.~27, no.~12, pp. 17\,298--17\,310, 2019.

\bibitem{Gnanasambandam2020}
A.~Gnanasambandam and S.~H. Chan, ``Hdr imaging with quanta image sensors:
  Theoretical limits and optimal reconstruction,'' \emph{IEEE Transactions on
  Computational Imaging}, vol.~6, pp. 1571--1585, 2020.

\bibitem{Elgendy2021}
O.~A. Elgendy, A.~Gnanasambandam, S.~H. Chan, and J.~Ma, ``Low-light
  demosaicking and denoising for small pixels using learned frequency
  selection,'' \emph{IEEE Transactions on Computational Imaging}, vol.~7, pp.
  137--150, 2021.

\bibitem{Buades2005}
A.~Buades, B.~Coll, and J.-M. Morel, ``A non-local algorithm for image
  denoising,'' in \emph{IEEE Conference on Computer Vision and Pattern
  Recognition (CVPR)}, 2005, pp. 60--65.

\bibitem{Dabov2007}
K.~Dabov, A.~Foi, V.~Katkovnik, and K.~Egiazarian, ``Image denoising by sparse
  3-d transform-domain collaborative filtering,'' \emph{IEEE Transactions on
  Image Processing}, vol.~16, no.~8, pp. 2080--2095, Aug. 2007.

\bibitem{Sun2014}
Z.~Sun, S.~Chen, and L.~Qiao, ``A general non-local denoising model using
  multi-kernel-induced measures,'' \emph{Pattern Recognition}, vol.~47, no.~4,
  pp. 1751--1763, 2014.

\bibitem{Li2016}
H.~Li and C.~Y. Suen, ``A novel non-local means image denoising method based on
  grey theory,'' \emph{Pattern Recognition}, vol.~49, pp. 237--248, 2016.

\bibitem{Elad2006}
M.~Elad and M.~Aharon, ``Image denoising via sparse and redundant
  representations over learned dictionaries,'' \emph{IEEE Transactions on Image
  Processing}, vol.~15, no.~12, pp. 3736--3745, 2006.

\bibitem{Mairal2009}
J.~Mairal, F.~Bach, J.~Ponce, G.~Sapiro, and A.~Zisserman, ``Non-local sparse
  models for image restoration,'' in \emph{IEEE International Conference on
  Computer Vision (ICCV)}, 2009, pp. 2272--2279.

\bibitem{Xu2018}
J.~Xu, L.~Zhang, and D.~Zhang, ``A trilateral weighted sparse coding scheme for
  real-world image denoising,'' in \emph{European Conference on Computer Vision
  (ECCV)}, 2018, pp. 20--36.

\bibitem{Zoran2011}
D.~Zoran and Y.~Weiss, ``From learning models of natural image patches to whole
  image restoration,'' in \emph{IEEE International Conference on Computer
  Vision (ICCV)}, 2011, pp. 479--486.

\bibitem{Chen2015}
F.~Chen, L.~Zhang, and H.~Yu, ``External patch prior guided internal clustering
  for image denoising,'' in \emph{IEEE International Conference on Computer
  Vision (ICCV)}, 2015, pp. 603--611.

\bibitem{Dong2013}
W.~Dong, G.~Shi, and X.~Li, ``Nonlocal image restoration with bilateral
  variance estimation: A low-rank approach,'' \emph{IEEE Transactions on Image
  Processing}, vol.~22, no.~2, pp. 700--711, 2013.

\bibitem{Gu2014}
S.~Gu, L.~Zhang, W.~Zuo, and X.~Feng, ``Weighted nuclear norm minimization with
  application to image denoising,'' in \emph{IEEE Conference on Computer Vision
  and Pattern Recognition (CVPR)}, 2014, pp. 2862--2869.

\bibitem{Chen2017}
Y.~Chen and T.~Pock, ``Trainable nonlinear reaction diffusion: A flexible
  framework for fast and effective image restoration,'' \emph{IEEE Transactions
  on Pattern Analysis and Machine Intelligence}, vol.~39, no.~6, pp.
  1256--1272, 2017.

\bibitem{Lefkimmiatis_2017}
S.~Lefkimmiatis, ``Non-local color image denoising with convolutional neural
  networks,'' in \emph{IEEE Conference on Computer Vision and Pattern
  Recognition (CVPR)}, 2017, pp. 3587--3596.

\bibitem{Zhang2017}
K.~Zhang, W.~Zuo, Y.~Chen, D.~Meng, and L.~Zhang, ``Beyond a gaussian denoiser:
  Residual learning of deep cnn for image denoising,'' \emph{IEEE Transactions
  on Image Processing}, vol.~26, no.~7, pp. 3142--3155, 2017.

\bibitem{Guo2019}
S.~Guo, Z.~Yan, K.~Zhang, W.~Zuo, and L.~Zhang, ``Toward convolutional blind
  denoising of real photographs,'' in \emph{IEEE Conference on Computer Vision
  and Pattern Recognition (CVPR)}, 2019, pp. 1712--1722.

\bibitem{Hong2019}
I.~Hong, Y.~Hwang, and D.~Kim, ``Efficient deep learning of image denoising
  using patch complexity local divide and deep conquer,'' \emph{Pattern
  Recognition}, vol.~96, p. 106945, 2019.

\bibitem{Anwar2019}
S.~Anwar and N.~Barnes, ``Real image denoising with feature attention,'' in
  \emph{IEEE/CVF International Conference on Computer Vision (ICCV)}, 2019, pp.
  3155--3164.

\bibitem{Tian2020}
C.~Tian, Y.~Xu, Z.~Li, W.~Zuo, L.~Fei, and H.~Liu, ``Attention-guided cnn for
  image denoising,'' \emph{Neural Networks}, vol. 124, pp. 117--129, 2020.

\bibitem{Zamir2020}
S.~W. Zamir, A.~Arora, S.~Khan, M.~Hayat, F.~S. Khan, M.-H. Yang, and L.~Shao,
  ``Learning enriched features for real image restoration and enhancement,'' in
  \emph{European Conference on Computer Vision (ECCV)}, 2020, pp. 492--511.

\bibitem{Lore2017}
K.~G. Lore, A.~Akintayo, and S.~Sarkar, ``Llnet: A deep autoencoder approach to
  natural low-light image enhancement,'' \emph{Pattern Recognition}, vol.~61,
  pp. 650--662, 2017.

\bibitem{Lv2018}
F.~Lv, F.~Lu, J.~Wu, and C.~Lim, ``{MBLLEN}: Low-light image/video enhancement
  using {CNNs},'' in \emph{British Machine Vision Conference (BMVC)}, 2018.

\bibitem{Jiang2021}
Y.~Jiang, X.~Gong, D.~Liu, Y.~Cheng, C.~Fang, X.~Shen, J.~Yang, P.~Zhou, and
  Z.~Wang, ``Enlightengan: Deep light enhancement without paired supervision,''
  \emph{IEEE Transactions on Image Processing}, vol.~30, pp. 2340--2349, 2021.

\bibitem{Wei2018}
C.~Wei, W.~Wang, W.~Yang, and J.~Liu, ``Deep retinex decomposition for
  low-light enhancement,'' in \emph{British Machine Vision Conference (BMVC)},
  2018.

\bibitem{Zhang2019}
Y.~Zhang, J.~Zhang, and X.~Guo, ``{Kindling the darkness}: A practical
  low-light image enhancer,'' in \emph{ACM International Conference on
  Multimedia}, 2019, pp. 1632--1640.

\bibitem{Zhang2021b}
S.~Zhang and E.~Y. Lam, ``An effective decomposition-enhancement method to
  restore light field images captured in the dark,'' \emph{Signal Processing},
  vol. 189, 2021.

\bibitem{Wu2018}
Y.~Wu and K.~He, ``{Group Normalization},'' in \emph{European Conference on
  Computer Vision (ECCV)}, 2018, pp. 3--19.

\bibitem{Wang2004}
Z.~Wang, A.~C. Bovik, H.~R. Sheikh, and E.~P. Simoncelli, ``Image quality
  assessment: from error visibility to structural similarity,'' \emph{IEEE
  Transactions on Image Processing}, vol.~13, no.~4, pp. 600--612, 2004.

\bibitem{Ledig2017}
C.~Ledig, L.~Theis, F.~Husz{\'{a}}r, J.~Caballero, A.~Cunningham, A.~Acosta,
  A.~Aitken, A.~Tejani, J.~Totz, Z.~Wang, and W.~Shi, ``Photo-realistic single
  image super-resolution using a generative adversarial network,'' in
  \emph{IEEE Conference on Computer Vision and Pattern Recognition (CVPR)},
  2017, pp. 4681--4690.

\bibitem{Zhang2021a}
S.~Zhang and E.~Y. Lam, ``Learning to restore light fields under low-light
  imaging,'' \emph{Neurocomputing}, vol. 456, pp. 76--87, 2021.

\bibitem{Abdelhamed2018}
A.~Abdelhamed, S.~Lin, and M.~S. Brown, ``A high-quality denoising dataset for
  smartphone cameras,'' in \emph{IEEE Conference on Computer Vision and Pattern
  Recognition (CVPR)}, 2018, pp. 1692--1700.

\end{thebibliography}

\end{document}